\documentclass[10pt,conference]{IEEEtran}

\IEEEoverridecommandlockouts
\usepackage{url}
\usepackage{cite}
\usepackage{soul}
\usepackage{float}
\usepackage[table]{xcolor}
\usepackage{balance}
\usepackage{amsmath}
\usepackage{amssymb}
\usepackage{siunitx}
\usepackage{graphicx}
\usepackage{textcomp}
\usepackage{stfloats}
\usepackage{verbatim}
\usepackage{multirow}
\usepackage{booktabs}
\usepackage{makecell}
\usepackage{colortbl}
\usepackage{enumitem}
\usepackage{amsfonts}
\usepackage{algorithmic}
\usepackage{fontawesome}
\usepackage[caption=false,font=normalsize,labelfont=sf,textfont=sf]{subfig}
\usepackage{threeparttable}
\usepackage{pifont}
\usepackage{adjustbox}
\usepackage{rotating}
\usepackage[hidelinks]{hyperref}
\usepackage{fancyhdr} 
\usepackage{inconsolata}
\usepackage{tcolorbox}
\usepackage[lined,boxed,linesnumbered,commentsnumbered,ruled]{algorithm2e}
\usepackage{color}
\usepackage{arydshln}
\usepackage{subcaption}
\usepackage[normalem]{ulem}

\definecolor{mygray}{gray}{.9} 
\NewDocumentCommand{\framecolorbox}{oommm}
 {
  \IfValueTF{#1}
   {\IfValueTF{#2}
    {\fcolorbox{#3}{#4}{\makebox[#1][#2]{#5}}}
    {\fcolorbox{#3}{#4}{\makebox[#1]{#5}}}%
   }
   {\fcolorbox{#3}{#4}{#5}}%
 }

\newcommand{\bench}{LLVM-Bench}
\newcommand{\tech}{LLVM-Ens}
\newcommand{\gym}{LLVM-Gym}

\newcommand{\Comment}[1]{}

\newtcolorbox{findings}{
  colback=gray!15,
  colframe=white,
  boxrule=0pt,
  sharp corners,
  left=5pt, right=5pt, top=5pt, bottom=5pt,
  before skip=5pt, after skip=5pt,
}

\def\BibTeX{{\rm B\kern-.05em{\sc i\kern-.025em b}\kern-.08em
    T\kern-.1667em\lower.7ex\hbox{E}\kern-.125emX}}

\newcommand{\distance}{4pt}
\begin{document}

\title{\bench{}: Benchmarking and Advancing Large Language Models for LLVM Compiler Issue Resolution}

\author{
    \IEEEauthorblockN{Zhao Tian}
    \IEEEauthorblockA{
    School of Computer Software, \\
    Tianjin University, Tianjin, China \\
    tianzhao@tju.edu.cn}
\and
\IEEEauthorblockN{\hphantom{Dummy}}
\IEEEauthorblockA{\hphantom{Dummy}}
\and
    \IEEEauthorblockN{Yingquan Zhao}
    \IEEEauthorblockA{
    School of Cybersecurity, \\
    Tianjin University, Tianjin, China \\
    zhaoyingquan@tju.edu.cn}
\and
\IEEEauthorblockN{\hphantom{Dummy}}
\IEEEauthorblockA{\hphantom{Dummy}}
\and
    \IEEEauthorblockN{Chenyao Suo}
    \IEEEauthorblockA{
    School of Computer Software, \\
    Tianjin University, Tianjin, China \\
    chenyaosuo@tju.edu.cn}
\and
\IEEEauthorblockN{\hphantom{Dummy}}
\IEEEauthorblockA{\hphantom{Dummy}}
\and
    \IEEEauthorblockN{Meng Wang}
    \IEEEauthorblockA{
    School of Computer Science, \\
    University of Bristol, Bristol, UK \\
    meng.wang@bristol.ac.uk}
\and
    \IEEEauthorblockN{Junjie Chen*}
    \IEEEauthorblockA{
    School of Computer Software, \\
    Tianjin University, Tianjin, China \\
    junjiechen@tju.edu.cn}
\and
\IEEEauthorblockN{\hphantom{Dummy}}
\IEEEauthorblockA{\hphantom{Dummy}}
\thanks{*Junjie Chen is the corresponding author.}
}

\maketitle

\thispagestyle{fancy}
\renewcommand{\headrulewidth}{0pt}
\renewcommand{\footrulewidth}{0pt} 
\pagestyle{fancy}

\begin{abstract}
LLVM is a widely used compiler infrastructure whose scale and complexity make issue resolution labor-intensive and challenging. 
Although large language models (LLMs) have recently achieved remarkable success in issue resolution, their effectiveness on complex system-level LLVM compiler remains largely unexplored.
To address this gap, we introduce \bench{}, the first large-scale benchmark for LLVM issue resolution, containing 423 real-world, validated tasks collected from the LLVM project. 
We further develop \gym{}, a scalable evaluation platform that automates issue reproduction, patch application, compiler building, and test execution.
Using \bench{} and \gym{}, we conduct a comprehensive study of four representative LLMs, six retrieval configurations, and three agents. 
Our results show that current LLM-based issue resolution techniques remain limited on \bench{}, with patch invalidity and build failures as the dominant failure modes. 
We further reveal a strong complementarity among different LLMs and agents, motivating \tech{}, a lightweight ensemble approach that expands the patch space through integrating the patches generated by diverse techniques, filters incorrect and redundant candidates, and identifies the most promising solution.
Our results show that \tech{} achieves a resolution rate of up to 21.99\%, further improving LLVM issue resolution.
\end{abstract}

\section{Introduction}
\label{sec:introduction}
LLVM compiler infrastructure is one of the most influential and widely adopted software systems in modern computing~\cite{lattner2004llvm,chen2020survey,zhou2021empirical}. 
It provides a collection of modular and reusable compiler and toolchain technologies that support a broad range of programming languages~\cite{sun2016toward,clang2026, flang2026, mlir2026}. 
Owing to its flexibility, extensibility, and high performance, LLVM has become a foundational component in both industrial and academic compiler and program analysis projects~\cite{lam2015numba,junod2015obfuscator,rust2026, retdec2026, DBLP:conf/osdi/CadarDE08, DBLP:conf/tacas/SchubertHB19}. 
As LLVM continues to evolve, however, developers must continuously address newly discovered bugs and feature requests. 
To date, LLVM community has accumulated nearly 100,000 issues on GitHub, of which approximately 30,000 remain unresolved~\cite{llvmissues2026}. 
Resolving these issues often requires understanding a large-scale and highly complex codebase, implementing non-trivial code changes, and preserving existing functionality while addressing the target issue~\cite{jimenezswe,li2025swe,wang2025solved}. 
Consequently, LLVM issue resolution remains labor-intensive and costly, motivating the development of automated issue resolution techniques to improve both developer productivity and software quality.

Recent advances in large language models (LLMs) have demonstrated remarkable capabilities in software issue resolution~\cite{yang2024swe,xia2024agentless,hou2024large}. 
Building upon the planning, reasoning, and tool-use abilities of LLMs, coding agents can autonomously navigate software repositories, execute commands, and generate patches for real-world issues. 
These capabilities have led to substantial progress on issue resolution benchmarks, most notably SWE-bench~\cite{jimenezswe}, a widely used benchmark that evaluates the ability of LLMs to resolve application-level software issues. 
Specifically, LLMs are provided only with an issue description and the corresponding codebase, and must autonomously generate a patch that passes all associated tests to resolve the issue~\cite{ruan2025specrover}. 
This task is highly challenging, requiring repository-level understanding, issue reproduction, fault localization, and precise code modification~\cite{yang2024swe,ma2025swe}. 
Nevertheless, with the rapid advancement of LLM-based issue resolution techniques, the resolution rate on SWE-bench-Verified has increased dramatically from 4.4\% when the benchmark was introduced in 2023 to nearly 95\% in recent state-of-the-art methods~\cite{valsai2026}.

Despite these advances, existing issue resolution benchmarks primarily focus on relatively small and less complex application-level software, such as Django~\cite{django2026} and Pytest~\cite{pytest2026}.
For example, the largest repository in SWE-bench contains approximately 6,000 files and 0.9 million lines of code, whereas LLVM comprises approximately 159,000 files and 53 million lines of code. 
Beyond its substantially larger scale, LLVM presents unique challenges stemming from intricate compiler architectures, complex program dependencies, and highly specialized domain knowledge. 
Given these characteristics and the practical importance of LLVM, an important question naturally arises: 
\ul{\textit{Can existing LLM-based issue resolution techniques effectively resolve complex real-world LLVM issues?}}
Answering this question requires both a large-scale benchmark of real-world LLVM issue resolution tasks and a systematic empirical evaluation of state-of-the-art LLM-based issue resolution techniques. 
However, existing application-level benchmarks do not include LLVM and therefore fail to capture its scale, complexity, and diversity. 
Consequently, the effectiveness of current issue resolution techniques on LLVM remains largely unexplored.

To bridge this gap, we construct \textbf{\bench{}}, the first large-scale benchmark for LLVM issue resolution, comprising 423 real-world tasks. 
Our data collection process begins with approximately 70,000 LLVM GitHub commits spanning 2023 to 2025. 
From these commits, we identify candidate tasks whose corresponding commits satisfy three criteria: (1) the commit is explicitly linked to a merged pull request, (2) it resolves a reported issue, and (3) it introduces verifiable test cases. 
To ensure reliable and reproducible evaluation, we develop \textbf{\gym{}}, a scalable and flexible platform that automates issue reproduction, patch application, compiler building, and LLVM test execution.
Candidate tasks that cannot be reliably reproduced or validated are discarded. 
To further ensure benchmark quality, we manually inspect all candidate tasks and remove those with ambiguous issue descriptions, incomplete reports, or potential solution leakage, following the official LLVM issue-reporting guidelines~\cite{llvmbugreport2026} and the validation principles adopted by prior work~\cite{swebenchvirified2024}. 
The resulting benchmark contains 423 high-quality tasks spanning four LLVM versions released over two years, three issue types (bug fix, optimization, and new feature), and all three major LLVM components (front-end, mid-end, and back-end).

To evaluate the effectiveness of state-of-the-art LLM-based issue resolution techniques on \bench{}, we conduct a comprehensive empirical study involving four representative LLMs (Gemini~\cite{team2023gemini}, Grok~\cite{grok2026}, DeepSeek~\cite{liu2024deepseek}, and Qwen~\cite{yang2025qwen3}). 
We evaluate these LLMs under six retrieval configurations 
and three agent frameworks (SWE-agent~\cite{yang2024swe}, Trae-agent~\cite{gao2025trae}, and Live-SWE-agent~\cite{xia2025live}), resulting in 36 experimental subjects. 
Our results show that retrieval-augmented LLMs achieve resolution rates below 5\% on \bench{}, while agent-based techniques improve performance but still achieve resolution rates below 11\%.
These results demonstrate that current issue resolution techniques remain far from effectively handling complex real-world LLVM issues.

To better understand these limitations, we further analyze the characteristics of successful and failed issue resolution attempts. 
We find that unresolved issues generally require larger and more complex patches and demand substantially greater human effort during their original resolution, suggesting that LLMs and human developers encounter similar challenges on difficult LLVM issues.
Moreover, the two dominant failure modes are patch invalidity, where generated patches cannot be applied to the LLVM codebase, and build failures, where applied patches prevent LLVM from compiling successfully. 
Since most failures occur before functional validation, these results suggest that current techniques struggle to reason effectively about repository structures, dependency relationships, and the constraints required to generate correct patches for large-scale LLVM systems.

Moreover, we observe strong complementarity among different LLMs and agents, with different techniques often resolving distinct subsets of issues. 
Motivated by this observation, we propose \tech{}, a lightweight ensemble approach that leverages the complementary strengths of diverse techniques by integrating their candidate patches, filtering incorrect and redundant candidates, and identifying the most promising solution.
Specifically, \tech{} filters out incorrect candidate patches using \gym{}, then removes redundant patches through code equivalence checking, and finally employs an LLM-based ensemble agent to identify the most promising solution. 
Experimental results show that \tech{} achieves a resolution rate of up to 21.99\%, substantially outperforming any individual technique.

The main contributions of this paper are as follows:
\begin{itemize}[leftmargin=10pt]
    \item \textbf{Benchmark}: We construct \bench{}, the first large-scale benchmark for real-world LLVM issue resolution. 
    It contains 423 high-quality, manually validated tasks, filling an important gap in the evaluation of LLM-based issue resolution techniques on the LLVM compiler.

    \item \textbf{Platform}: We develop \gym{}, a scalable and flexible evaluation platform that automates issue reproduction, patch application, compiler building, and LLVM test execution, enabling reproducible large-scale evaluation.
    
    \item \textbf{Study}: We conduct the first comprehensive empirical study of LLM-based LLVM issue resolution, evaluating four LLMs, six retrieval configurations, and three agents (36 experimental subjects). 
    Our results reveal that existing techniques remain limited on complex LLVM issues, with patch invalidity and build failures as the dominant failure modes, suggesting challenges in modeling LLVM repository structures, dependencies, and and constraints.

    \item \textbf{Technique}: Inspired by complementarity among different LLMs and agents, we propose \tech{}, a lightweight ensemble approach that expands the patch space through integrating the patches generated by diverse techniques, filters incorrect and redundant candidates, and identifies the most promising solution.
    Experimental results show that \tech{} further improves LLVM issue resolution, outperforming each individual technique.

\end{itemize}

\section{\bench{}}
\label{sec:benchmark}

\subsection{Benchmark Construction}
\label{subsec:construction}
\begin{figure}[t]
    \centering
    \includegraphics[width=1.0\linewidth]{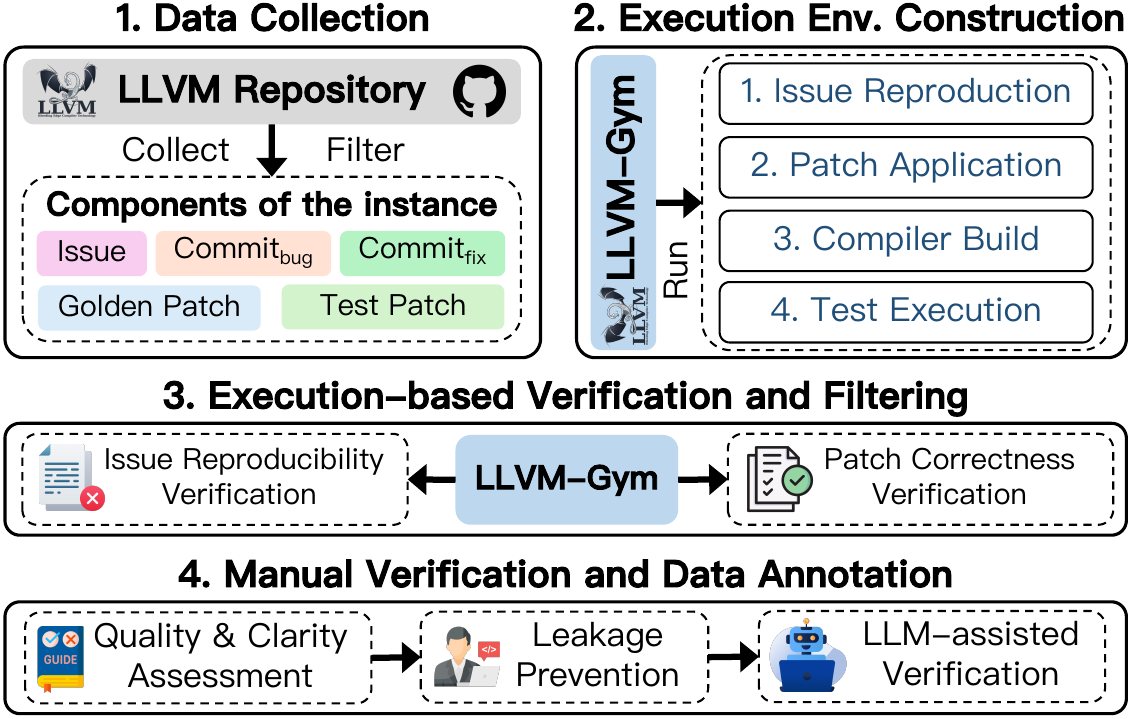}
    \caption{The pipeline of benchmark construction}
    \label{fig:overview}
\end{figure}

Figure~\ref{fig:overview} illustrates the construction pipeline of \bench{}, which is described as follows.

\subsubsection{Data Collection}
\label{subsubsec:data_collection}
To construct a high-quality benchmark for our study, we systematically crawl the commit history of the LLVM repository, initially retrieving approximately 561,000 GitHub commits. 
Our study focuses on four major LLVM versions (18 -- 21) released over two years.
After preliminary filtering, the pool is narrowed to approximately 70,000 commits.
Specifically, we identify pairs of consecutive commits that collectively represent the resolution of a single LLVM issue. 
Within each pair, the preceding commit is denoted as $Commit_{bug}$ and the succeeding one as $Commit_{fix}$. 
In addition, we decompose the code changes in these commits into two distinct components: 
(1) \textit{Test Patch}: defined as the code diff in test files between $Commit_{bug}$ and $Commit_{fix}$, including all newly added or modified test cases.
The test patch encapsulates the complete conditions required to issue reproduction and patch verification, such as the target compiler component, test programs, compiler options, and other relevant execution settings.
(2) \textit{Golden Patch}: defined as the remaining code diff (excluding the test patch) that directly resolves the issue.
To ensure data integrity, each instance must be explicitly linked to a reported issue and a merged pull request. 
Furthermore, as LLVM is primarily developed in C/C++, we restrict our collection to modifications involving core source files (i.e., \textit{.c}, \textit{.h}, \textit{.cpp}, \textit{.hpp}, and \textit{.ll}).
We also filter out ``meta-changes'' that merely involve renaming files or directories. 
Following this rigorous collection process, we obtain a set of 1,222 candidate instances.

\subsubsection{Execution Environment Construction}
\label{subsubsec:llvm_gym}
To ensure reliable issue reproduction and rigorous patch verification, we develop \textbf{\gym{}}, a scalable evaluation platform that provides the following four core capabilities: 
(1) \textit{Issue Reproduction}: leverages the test patch to trigger or confirm the specific compiler issues.
(2) \textit{Patch Application}: applies the candidate patch to the corresponding version of the LLVM codebase.
(3) \textit{Compiler Building}: builds the patched LLVM codebase into executable compiler binaries using the \texttt{Ninja} build system~\cite{ninja2026}.
(4) \textit{Test Execution}: executes the complete LLVM test suite, including both existing regression tests and newly added tests in the test patch, to verify that the patch resolves the issue without introducing regressions.
Therefore, \gym{} streamlines large-scale evaluation by automating the entire workflow.

\subsubsection{Execution-based Verification and Filtering}
\label{subsubsec:execution_based_filtering}
Leveraging the \gym{}, we conduct a multi-stage execution-based verification process on the 1,222 candidate instances. 
This step ensures that each issue is reliably reproducible and that both the golden patch and test patch are consistently correct.
It consists of the following two verification stages: 
(1) \textit{Issue Reproducibility Verification}: To confirm the presence of the reported issue, \gym{} applies and executes the test patch to $Commit_{bug}$. 
An instance is retained only if the test patch fails as expected, thereby successfully triggering or confirming the reported issue.
(2) \textit{Patch Correctness Verification}: To ensure the correctness of the golden patch, \gym{} applies both the golden patch and the test patch to $Commit_{bug}$, and executes the entire test suite, including all existing regression tests (in the original codebase) and the new tests (in the test patch). 
An instance is considered correct only if all tests pass, demonstrating that the golden patch not only resolves the specific issue but also introduces no regressions.
To further guarantee the robustness of our benchmark, multiple authors independently perform this automated verification pipeline.
Any instance failing to meet these stringent criteria is purged, yielding a refined benchmark of 492 candidate instances.

\subsubsection{Manual Verification and Data Annotation}
\label{subsubsec:manual_verification}
Following existing work~\cite{jimenezswe,xiang2026evaluating}, we conduct a rigorous manual inspection of the candidate instances to further ensure their quality and integrity. 
This manual verification process is executed in two primary stages:
(1) \textit{Quality and Clarity Assessment}: Adhering to the official LLVM issue reporting guidelines~\cite{llvmbugreport2026}, multiple authors independently review each reported issue. 
We ensure that the issue descriptions are self-contained, unambiguous, and provide sufficient context for the developers to understand the underlying technical challenge.
(2) \textit{Leakage Prevention}: To safeguard against potential data contamination or solution leakage, we meticulously scrutinize each issue description. 
Any instance containing explicit or implicit hints of the final solution (e.g., code snippets about the fix or direct pointers to specific logic changes) is promptly discarded.
In addition, each verified instance is categorized based on its functional nature into one of three primary issue types (i.e., bug fix, optimization, or new feature).
To further bolster the reliability of our manual verification, we employ an advanced LLM (i.e., Gemini-3) to perform a secondary, automated audit of the verification results, serving as a redundant check for consistency. 
Finally, we obtain the \bench{} comprising 423 high-quality instances.

\begin{table}[t!]
    \caption{Task scale and complexity of \bench{}.}
    \label{tab:statistics}
    \centering
    \tabcolsep=2.03mm
    \small
    \begin{tabular}{llccc}
        \toprule
        \textbf{Category} & \textbf{Metric} & \textbf{Mean} & \textbf{Median} & \textbf{Max} \\ \midrule
        \textbf{Issue Text} & \#Tokens & 1,323.67 & 457 & 21,580 \\ 
        \midrule
        \multirow{2}{*}{\textbf{Codebase}} 
        & \#Files & 145.49K & 145.29K & 159.92K \\
        & \#Lines & 47.95M & 47.78M & 53.23M \\
        \midrule
        \multirow{3}{*}{\textbf{Golden Patch}} 
        & \#Files & 3.36 & 2 & 53 \\
        & \#Lines & 129.97 & 52 & 3,775 \\
        & \#Tokens & 698.01 & 369 & 13,563 \\ 
        \midrule
        \multirow{2}{*}{\textbf{\makecell[l]{Manual\\Resolution}}} 
        & Time (day) & 72.46 & 5.99 & 4,016.95 \\
        & \#Discussion & 3.70 & 3 & 45 \\
        \bottomrule
    \end{tabular}
\end{table}

\subsection{Characteristics of \bench{}}
\label{subsec:characteristics}
Table~\ref{tab:statistics} presents the task scale and complexity of \bench{}. 
First, the issue descriptions in \bench{} are highly detailed, with average, median, and maximum token lengths of 1,323.67, 457, and 21,580, respectively.
A defining characteristic of \bench{} is the large scale of codebase. 
On average, each task is associated with a codebase version containing 149.49K files and 47.95M lines of code (LOC), while the latest LLVM version contains up to 159.92K files and 53.23M LOC. 
Beyond codebase scale, \bench{} also exhibits substantially higher patch complexity. 
Resolving a single issue requires modifications to an average of 3.36 files, involving 129.97 lines and 698.01 tokens.

Moreover, the original human resolution process for these issues is highly non-trivial. 
On average, each issue requires 72.46 days from creation to pull request merge and involves 3.7 rounds of developer discussion. 
We further analyze the distribution of resolution durations and observe that 25.53\% of issues are resolved within one day, 27.19\% within one week, 21.04\% within one month, and 18.68\% within six months, while 7.57\% require more than six months. 
These results indicate that \bench{} captures a diverse spectrum of real-world LLVM issue resolution scenarios, ranging from rapid fixes to long-standing and highly complex issues.

To further characterize \bench{}, we manually categorize all 423 task instances according to three dimensions (i.e., issue type, LLVM component, and LLVM version). 
Table~\ref{tab:distribution} summarizes the detailed task distributions. 
Regarding issue type, \bench{} consists of bug fixes, optimizations, and new features, accounting for 75.41\%, 19.39\%, and 5.20\%, respectively, indicating that bug fix is the dominant category in LLVM issues.
Following prior work~\cite{lattner2004llvm,sun2016toward}, LLVM is organized into three major components (i.e., front-end, mid-end, and back-end), each responsible for different stages of compiler functionality. 
Table~\ref{tab:distribution} shows that \bench{} includes issues spanning all three major components. 
Specifically, front-end, mid-end, back-end, and others account for 34.04\%, 58.16\%, 6.62\%, and 1.18\% of \bench{}, respectively, suggesting that mid-end issues constitute the largest proportion of LLVM issue resolution tasks.
In addition, LLVM evolves rapidly through continuous contributions from thousands of developers worldwide, with a new major release approximately every six months and long-term maintenance support provided for each version. 
As shown in Table~\ref{tab:distribution}, \bench{} covers four major versions (versions 18--21) released over the past two years.
Capturing such version diversity is essential for constructing a realistic benchmark.

\begin{table}[t!]
    \caption{Task distribution of \bench{}.}
    \label{tab:distribution}
    \centering
    \tabcolsep=2.5mm
    \small
    \begin{tabular}{llc}
        \toprule
        \multicolumn{2}{c}{\textbf{Task Category}} & \textbf{\#Tasks (\%Percentage)} \\ \midrule
        \multirow{3}{*}{\textbf{Issue Type}} 
        & Bug Fix & 319 (75.41\%) \\
        & Optimization & 82 (19.39\%) \\
        & New Feature & 22 (5.20\%) \\ \midrule
        \multirow{4}{*}{\textbf{LLVM Component}} 
        & Front-end & 144 (34.04\%) \\
        & Mid-end & 246 (58.16\%) \\
        & Back-end & 28 (6.62\%) \\ 
        & Others & 5 (1.18\%) \\ \midrule
        \multirow{4}{*}{\textbf{LLVM Version}} 
        & 18.x.x & 111 (26.24\%) \\
        & 19.x.x & 121 (28.61\%) \\
        & 20.x.x & 99 (23.40\%) \\ 
        & 21.x.x & 92 (21.75\%) \\ 
        \bottomrule
    \end{tabular}
\end{table}
\section{Study Design}
\label{sec:evaluation_design}
\subsection{Research Questions}
\label{subsec:rq}
Our study aims to address the following research questions:
\begin{itemize}[leftmargin=10pt]
    \item \textbf{RQ1}: How do different LLMs perform on \bench{}?
    
    \item \textbf{RQ2}: How do different agents perform on \bench{}?
    
    \item \textbf{RQ3}: What are the error characteristics of the studied issue resolution techniques on \bench{}?
    
    \item \textbf{RQ4}: How does \tech{} perform on \bench{}?
\end{itemize}


\subsection{Experimented LLMs}
\label{subsec:llm}
In our experiments, we evaluate four advanced open-source and commercial LLMs. 
For open-source LLMs, we select two representative LLMs: DeepSeek (version \texttt{deepseek-v3.2}) and Qwen (version \texttt{qwen3-coder-plus}). 
For commercial LLMs, we include two widely-adopted advanced LLMs: Gemini (version \texttt{gemini-3-flash}) and Grok (version \texttt{grok-code-fast-1}).
These LLMs have demonstrated strong capabilities in many software engineering tasks, and are broadly adopted in both academia and industry~\cite{xia2024agentless,gao2025trae,wang2025openhands}. 
In addition, we set the temperature to 0 to minimize output randomness and improve experimental stability. 
Besides, we configure the maximum generation length to 8,192.

\subsection{Issue Resolution Techniques}
\label{subsec:baseline}

In our experiments, we evaluate five representative or state-of-the-art issue resolution techniques:
\begin{itemize}[leftmargin=10pt]
    \item \textbf{Sparse Retrieval}~\cite{jimenezswe}: 
    It is a retrieval-augmented approach that employs the BM25~\cite{robertson1994some} algorithm to retrieve issue-relevant information from the codebase for LLM-based localization and patch generation.
    
    \item \textbf{Oracle Retrieval}~\cite{jimenezswe}: 
    It retrieves contextual information directly from the files modified by the gold patch. 
    Although unavailable in practical issue resolution, this setting serves as an approximate upper bound for retrieval quality and helps isolate the impact of retrieval on downstream issue resolution performance.
    
    \item \textbf{SWE-agent}~\cite{yang2024swe}: 
    It designs an agent-computer interface that enables agents to interact with the repository environment through executing bash commands and navigating the codebase. 
    Through this interactive framework, SWE-agent can autonomously address real-world software issues.

    \item \textbf{Trae-agent}~\cite{gao2025trae}: It is an industrial agent designed for software issue resolution tasks. 
    It provides a command-line interface that can accurately interpret natural language instructions and autonomously execute complex workflows through a diverse set of sophisticated agent tools.

    \item \textbf{Live-SWE-agent}~\cite{xia2025live}: It is a novel self-evolving agent that can dynamically evolve its agent harness without requiring offline training or additional optimization pipelines for improving issue resolution effectiveness. 
    
\end{itemize}

Note that both retrieval-augmented techniques (Sparse Retrieval and Oracle Retrieval) primarily serve to provide relevant contextual information to LLMs, thereby enabling us to evaluate the intrinsic issue resolution capabilities of different LLMs (as shown in \textbf{RQ1}).
Following prior work~\cite{jimenezswe}, we evaluate three context-length configurations (13K, 20K, and 50K) for two techniques.
Considering the trade-off between effectiveness and computational cost, we limit the maximum interaction turns to 50 for all agents.
Overall, our experiments include 36 baselines in total (9 techniques $\times$ 4 LLMs).

\subsection{Evaluation Metrics}
\label{subsec:metric}
Following prior work~\cite{jimenezswe,mathai2024kgym,xiang2026evaluating}, we evaluate the effectiveness of issue resolution techniques using \textit{\%Applied}, \textit{\%Built}, and \textit{\%Resolved}. 
In addition, we measure their efficiency using \textit{\#Input Tokens}, \textit{\#Output Tokens}, and \textit{\$Cost}.

\begin{itemize}[leftmargin=10pt]
    \item \ul{\textbf{\textit{\%Applied}}} measures the syntactic correctness of patches by determining whether a patch can be successfully applied to the LLVM codebase. 
    It is computed as the percentage of successfully applied patches among all generated patches.

    \item \ul{\textbf{\textit{\%Built}}} measures the compilation correctness of patches by determining whether the LLVM codebase containing the generated patch can be successfully built. 
    It is calculated as the percentage of successfully built patches among all generated patches.

    \item \ul{\textbf{\textit{\%Resolved}}} measures the functional correctness of patches by assessing whether they successfully resolve the target LLVM issues. 
    It is defined as the percentage of successfully resolved issues out of the total number of evaluated issues.

    \item \ul{\textbf{\textit{\#Input Tokens}}} and \ul{\textbf{\textit{\#Output Tokens}}} denote the average numbers of input and output tokens consumed by the LLM, respectively.
    These metrics quantify the token-level computational overhead of each technique in practical deployment.
    
    \item \ul{\textbf{\textit{\$Cost}}} denotes the average monetary cost of LLM inference. 
    
\end{itemize}

\section{Results and Analysis}
\label{sec:results_and_analysis}

\begin{table*}[t!]
    \caption{Effectiveness and efficiency comparison of different LLMs and retrieval strategies on \bench{}.}
    \label{tab:rq1_main}
    \centering
    \tabcolsep=1.70mm
    \small
    \begin{tabular}{lcccccccccccc}
        \toprule
        \multirow{2}{*}{\textbf{Metrics}} & \multicolumn{3}{c}{\textbf{DeepSeek}} & \multicolumn{3}{c}{\textbf{Qwen}} & \multicolumn{3}{c}{\textbf{Gemini}} & \multicolumn{3}{c}{\textbf{Grok}} \\ \cmidrule(lr){2-4} \cmidrule(lr){5-7} \cmidrule(lr){8-10} \cmidrule(lr){11-13}
        & w/ \textbf{13K} & w/ \textbf{27K} & w/ \textbf{50K} & w/ \textbf{13K} & w/ \textbf{27K} & w/ \textbf{50K} & w/ \textbf{13K} & w/ \textbf{27K} & w/ \textbf{50K} & w/ \textbf{13K} & w/ \textbf{27K} & w/ \textbf{50K}  \\
        \midrule
        \multicolumn{13}{l}{\framecolorbox[17.55cm][l]{gray!30}{gray!30}{\textbf{Sparse Retrieval}}} \\[3pt] \hdashline \\[-7pt]
        \textbf{\%Applied} \textcolor{green}{$\uparrow$} & 22.46\% & 24.82\% & 25.06\% & 30.50\% & 27.66\% & 28.61\% & 39.95\% & \cellcolor{green!20}\textbf{41.61\%} & 41.37\% & 19.39\% & 17.49\% & 20.33\% \\
        \textbf{\%Built} \textcolor{green}{$\uparrow$} & 4.49\% & 6.86\% & 6.86\% & 15.84\% & 12.06\% & 15.84\% & 9.69\% & 12.53\% & \cellcolor{green!20}\textbf{18.44\%} & 3.78\% & 2.36\% & 3.07\% \\
        \textbf{\%Resolved} \textcolor{green}{$\uparrow$} & 0.71\% & 1.18\% & 1.42\% & 0.47\% & 0.24\% & 0.71\% & 1.18\% & 1.42\% & 1.65\% & 0.71\% & 0.95\% & \cellcolor{green!20}\textbf{1.89\%} \\
        \textbf{\#Input} \textcolor{red}{$\downarrow$} & 15,625 & 30,448 & 54,820 & 15,209 & 29,587 & 53,099 & 17,124 & 33,245 & 59,648 & \cellcolor{red!20}\textbf{14,625} & 28,453 & 51,166 \\
        \textbf{\#Output} \textcolor{red}{$\downarrow$} & 3,658 & 3,957 & 3,645 & 1,212 & 1,351 & 1,241 & 390 & \cellcolor{red!20}\textbf{382} & 432 & 7,035 & 6,636 & 6,785 \\
        \textbf{\$Cost} \textcolor{red}{$\downarrow$} & \cellcolor{red!20}\textbf{0.003} & 0.004 & 0.006 & 0.035 & 0.061 & 0.101 & 0.010 & 0.018 & 0.031 & 0.012 & 0.012 & 0.014 \\
        \midrule
        \multicolumn{13}{l}{\framecolorbox[17.55cm][l]{gray!30}{gray!30}{\textbf{Oracle Retrieval}}} \\[2pt] \hdashline \\[-7pt]
        \textbf{\%Applied} \textcolor{green}{$\uparrow$} & 32.39\% & 34.52\% & 35.93\% & 40.90\% & 41.61\% & 44.92\% & 52.48\% & 47.75\% & \cellcolor{green!20}\textbf{53.66\%} & 24.11\% & 23.88\% & 23.88\% \\
        \textbf{\%Built} \textcolor{green}{$\uparrow$} & 15.60\% & 18.44\% & 21.51\% & 22.22\% & 22.93\% & 22.70\% & 26.24\% & 29.31\% & \cellcolor{green!20}\textbf{35.22\%} & 11.35\% & 6.38\% & 4.73\% \\
        \textbf{\%Resolved} \textcolor{green}{$\uparrow$} & 2.84\% & 3.07\% & 3.55\% & 1.42\% & 1.42\% & 1.65\% & 2.60\% & 3.07\% & 3.55\% & 3.55\% & 2.60\% & \cellcolor{green!20}\textbf{4.02\%} \\
        \textbf{\#Input} \textcolor{red}{$\downarrow$} & 14,476 & 25,628 & 39,247 & 13,989 & 24,520 & 37,634 & 15,575 & 27,445 & 42,118 & \cellcolor{red!20}\textbf{13,602} & 24,070 & 36,986 \\
        \textbf{\#Output} \textcolor{red}{$\downarrow$} & 3,438 & 3,887 & 3,689 & 708 & 887 & 725 & 401 & \cellcolor{red!20}\textbf{399} & 410 & 6,857 & 6,690 & 7,003 \\
        \textbf{\$Cost} \textcolor{red}{$\downarrow$} & \cellcolor{red!20}\textbf{0.003} & 0.004 & 0.005 & 0.029 & 0.049 & 0.071 & 0.009 & 0.015 & 0.022 & 0.011 & 0.012 & 0.013 \\
        \bottomrule
    \end{tabular}
\end{table*}

\subsection{RQ1: Performance of LLMs on \bench{}}
\label{subsec:RQ1}

Table~\ref{tab:rq1_main} presents the effectiveness and efficiency comparison results of different LLMs on \bench{}.

\textbf{Comparison among LLMs.}
In effectiveness evaluation, we first observe that Grok achieves the highest average \textit{\%Resolved} score (2.29\%), slightly outperforming Gemini (2.25\%) and DeepSeek (2.13\%), and substantially surpassing Qwen (0.99\%). 
In addition, Gemini achieves the highest average \textit{\%Built} (21.91\%) and \textit{\%Applied} (46.14\%) scores among all evaluated LLMs.
In efficiency evaluation, we observe that all LLMs consume nearly identical numbers of \textit{\#Input Tokens}, while their \textit{\#Output Tokens} differ substantially. 
Specifically, DeepSeek, Qwen, Gemini, and Grok generate an average of 3,712, 1,021, 402, and 6,834 output tokens, respectively. 
Since Grok achieves the highest \textit{\%Resolved} score, we further manually inspect its generated outputs and find that it tends to produce substantially more reasoning and deliberation during patch generation, thereby resulting in higher output token consumption. 
This observation suggests that richer reasoning processes may contribute to resolving more challenging LLVM issues.
Besides, DeepSeek incurs the lowest average monetary cost (\$0.004), whereas Qwen incurs the highest (\$0.058). 

\begin{findings}
\textbf{\textit{Finding 1:}} 
All evaluated LLMs achieve \textit{\%Resolved} scores below 5\%, indicating that LLMs still exhibit limited capability in handling complex LLVM issues.
\end{findings}

\textbf{Comparison among Retrieval-augmented Techniques.}
In effectiveness evaluation, we observe that LLMs equipped with Oracle Retrieval achieve average scores of 2.78\%, 19.72\%, and 38.00\% in the \textit{\%Resolved}, \textit{\%Built}, and \textit{\%Applied} metrics, respectively. 
These results consistently outperform LLMs equipped with Sparse Retrieval, which achieve 1.04\%, 9.32\%, and 28.27\% on the corresponding metrics, yielding substantial relative improvements of 167.31\%, 111.59\%, and 34.42\%, respectively. 
These findings indicate that providing contextual information from the files modified by the golden patch can substantially improve LLVM issue resolution.
However, despite the advantage of Oracle Retrieval, its best-performing \textit{\%Resolved} score reaches only 4.02\%, indicating that LLVM issue resolution tasks remain highly challenging even under an optimistic and potentially upper-bound retrieval setting.
It suggests that LLVM issue resolution requires context beyond isolated files, incorporating broader compiler structures, dependencies, and constraints.
Furthermore, since both retrieval-augmented techniques share identical experimental configurations (i.e., base LLMs and context-length settings), their efficiency metrics exhibit no substantial differences.

\begin{findings}
\textbf{\textit{Finding 2:}} 
Oracle Retrieval consistently outperforms Sparse Retrieval by leveraging context from the golden-patch files.
However, its best \textit{\%Resolved} reaches only 4.02\%, indicating that LLVM issue resolution requires context beyond the isolated files, such as compiler structures, dependencies, and constraints.

\end{findings}

\textbf{Comparison among Context-length Configurations.}
We observe that the performance of LLMs generally improves as the context length increases. 
Specifically, the average \textit{\%Resolved} scores under the 13K, 27K, and 50K context-length configurations are 1.69\%, 1.74\%, and 2.31\%, respectively, demonstrating a gradual performance improvement with larger contexts.
However, this improvement is accompanied by increased computational overhead. 
In particular, \textit{\#Input Tokens}, \textit{\#Output Tokens}, and \textit{\$Cost} all increase consistently as the context length grows. 
These findings suggest that larger contexts may provide more useful repository information, thereby helping LLMs better understand and resolve complex LLVM issues. 
At the same time, the trade-off between effectiveness and efficiency must be carefully considered in practical deployment scenarios.
In future work, we plan to further investigate the potential benefits and scalability of larger context windows for LLVM issue resolution.

\begin{findings}
\textbf{\textit{Finding 3:}} 
Increasing the retrieval context from 13K to 50K yields an average 36.69\% relative improvement in \textit{\%Resolved}, albeit with higher computational cost.
\end{findings}

\begin{figure*}[t!]
    \centering
    \includegraphics[width=1.0\linewidth]{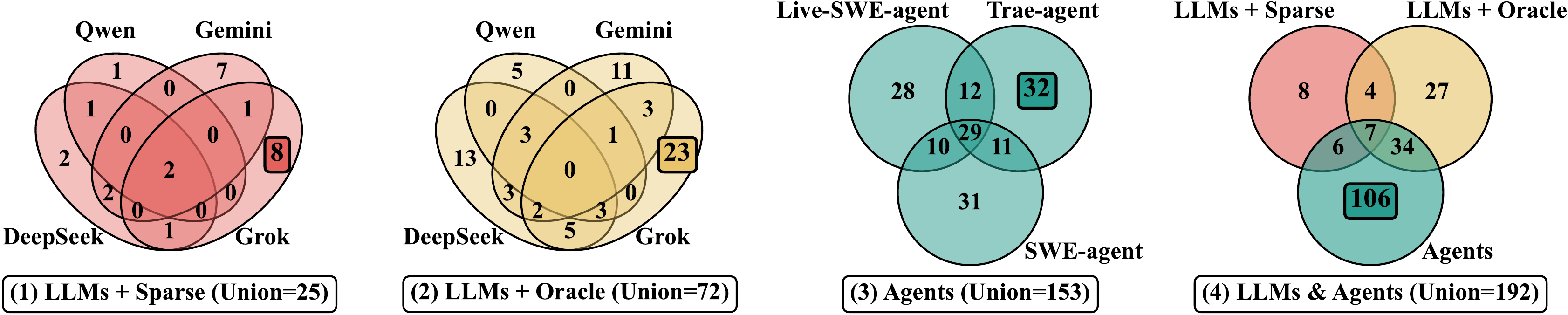}
    \caption{Number of uniquely resolved issues (\textcolor{green}{$\uparrow$}) across different LLMs and agents on \bench{}}
    \label{fig:rq2_venn}
\end{figure*}

\textbf{Complementarity of Different LLMs.}
Figures~\ref{fig:rq2_venn}(1) and ~\ref{fig:rq2_venn}(2) present the number of uniquely resolved issues achieved by different LLMs on \bench{}. 
We observe a notable degree of complementarity among the evaluated LLMs.
Specifically, under the Sparse Retrieval setting, the individual \textit{\%Resolved} scores of different LLMs range from 0.24\% to 1.89\%, whereas the union of all resolved issues across LLMs reaches 5.91\%. 
Similarly, under the Oracle Retrieval setting, the individual \textit{\%Resolved} scores range from 1.42\% to 4.02\%, while the union of resolved LLVM issues increases substantially to 17.02\%.
These findings indicate that, although individual LLMs achieve relatively low performance on \bench{}, different LLMs tend to resolve different subsets of issues, demonstrating strong complementary capabilities. 

\begin{findings}
\textbf{\textit{Finding 4:}} 
Different LLMs exhibit strong complementarity, with each model resolving distinct subsets of issues.
\end{findings}

\begin{table*}[t!]
    \caption{Effectiveness and efficiency comparison of different agents on \bench{}.}
    \label{tab:rq2_main}
    \centering
    \tabcolsep=1.23mm
    \small
    \begin{tabular}{lcccccccccccc}
        \toprule
        \multirow{2}{*}{\textbf{Metrics}} & \multicolumn{4}{c}{\textbf{SWE-agent}} & \multicolumn{4}{c}{\textbf{Trae-agent}} & \multicolumn{4}{c}{\textbf{Live-SWE-agent}} \\ \cmidrule(lr){2-5} \cmidrule(lr){6-9} \cmidrule(lr){10-13}
        & DeepSeek & Qwen & Gemini & Grok & DeepSeek & Qwen & Gemini & Grok & DeepSeek & Qwen & Gemini & Grok  \\
        \midrule
        \textbf{\%Applied} \textcolor{green}{$\uparrow$} & \cellcolor{green!20}\textbf{55.32\%} & 27.90\% & 30.97\% & 43.26\% & 48.23\% & 21.75\% & 41.13\% & 29.55\% & 46.81\% & 23.64\% & 30.97\% & 43.97\% \\
        \textbf{\%Built} \textcolor{green}{$\uparrow$} & \cellcolor{green!20}\textbf{40.19\%} & 20.33\% & 23.40\% & 17.97\% & 36.88\% & 18.91\% & 19.86\% & 12.06\% & 35.70\% & 17.02\% & 6.86\% & 19.15\% \\
        \textbf{\%Resolved} \textcolor{green}{$\uparrow$} & 7.80\% & 3.78\% & 4.26\% & \cellcolor{green!20}\textbf{10.87\%} & 8.98\% & 3.55\% & 4.49\% & 7.33\% & 8.51\% & 2.60\% & 4.73\% & 7.57\% \\ 
        \textbf{\#Input} \textcolor{red}{$\downarrow$} & 314,562 & 247,208 & 201,168 & 474,261 & 270,279 & 223,002 & 199,130 & 307,315 & 382,775 & 309,109 & \cellcolor{red!20}\textbf{191,273} & 500,309 \\
        \textbf{\#Output} \textcolor{red}{$\downarrow$} & 6,526 & 8,080 & 15,020 & 21,003 & 6,772 & 18,314 & \cellcolor{red!20}\textbf{5,819} & 34,802 & 7,383 & 9,182 & 13,756 & 22,256 \\
        \textbf{\$Cost} \textcolor{red}{$\downarrow$} & 0.027 & 0.486 & 0.146 & 0.068 & \cellcolor{red!20}\textbf{0.024} & 0.515 & 0.117 & 0.076 & 0.033 & 0.601 & 0.137 & 0.072 \\
        \bottomrule
    \end{tabular}
\end{table*}

\subsection{RQ2: Performance of Agents on \bench{}}
\label{subsec:RQ2}


Table~\ref{tab:rq2_main} shows the effectiveness and efficiency comparison results of different agents on \bench{}.

\textbf{Comparison with LLM-based Techniques.}
RQ1 and RQ2 separately evaluate the performance of LLM- and agent-based techniques on \bench{}. 
We compare these two categories of approaches to better understand the impact of agent frameworks on LLVM issue resolution.
In effectiveness evaluation, agents achieve average scores of 6.21\%, 22.36\%, and 36.96\% in terms of \textit{\%Resolved}, \textit{\%Built}, and \textit{\%Applied} metrics, respectively, consistently outperforming LLM-based techniques, which achieve 1.91\%, 14.52\%, and 33.14\% on the corresponding metrics. 
These improvements correspond to relative gains of 206.21\%, 35.57\%, and 9.69\%, respectively. 
The results demonstrate that well-designed agent frameworks can substantially enhance the LLVM issue resolution capability of LLMs.
In efficiency evaluation, however, all agents incur significantly higher computational overhead. 
Specifically, agents consume an average of 293,366, 14,076, and \$0.188 in terms of \textit{\#Input Tokens}, \textit{\#Output Tokens}, and \textit{\$Cost}, respectively, compared with 29,931, 2,992, and \$0.023 for LLM-based techniques. 
These differences correspond to relative increases of 880.14\%, 370.45\%, and 717.39\%, respectively.
These findings suggest that the additional interactions introduced by agent frameworks, including iterative reasoning, task orchestration, environment interaction, and tool use, substantially increase token consumption and inference costs, while simultaneously improving LLVM issue resolution performance. 
Overall, these results indicate that advanced agent frameworks are beneficial for resolving complex LLVM issues, although the trade-off between effectiveness and computational costs must be carefully considered. 

\begin{findings}
\textbf{\textit{Finding 5:}} 
Agent-based techniques consistently outperform LLM-based techniques, with an average 206.21\% relative improvement in \textit{\%Resolved} metric, while incurring substantially higher computational costs.
\end{findings}

\textbf{Comparison among Agent Frameworks.}
In effectiveness evaluation, we first observe that SWE-agent achieves the highest average \textit{\%Resolved} score (6.68\%), outperforming Trae-agent (6.09\%) and Live-SWE-agent (5.85\%). 
In addition, SWE-agent consistently achieves the highest average \textit{\%Built} (25.47\%) and \textit{\%Applied} (39.36\%) scores among all evaluated agents.
In efficiency evaluation, we observe that Live-SWE-agent incurs the highest average \textit{\#Input Tokens} (345,866) and \textit{\$Cost} (\$0.211), whereas Trae-agent generates the largest number of \textit{\#Output Tokens} (16,427). 
These results indicate that SWE-agent achieves a better balance between effectiveness and efficiency compared with the other evaluated agent frameworks.
However, across all experimental settings, the \textit{\%Resolved} scores of the studied agents remain below 11\%, suggesting that even advanced agent-based techniques still exhibit limited capability in LLVM issue resolution.

\begin{findings}
\textbf{\textit{Finding 6:}} 
All evaluated agents achieve \textit{\%Resolved} scores below 11\%, suggesting that current agent-based techniques remain limited in resolving complex LLVM issues.
\end{findings}

\textbf{Complementarity of Different Agents.}
Figure~\ref{fig:rq2_venn}(3) shows the number of uniquely resolved issues achieved by different agents on \bench{}. 
We observe substantial complementarity among the evaluated agent frameworks. Specifically, although the individual \textit{\%Resolved} scores of SWE-agent, Trae-agent, and Live-SWE-agent range from only 2.60\% to 10.87\%, the union of resolved issues across all agents reaches 36.17\%.
Figure~\ref{fig:rq2_venn}(4) further extends this analysis by combining the results of agents with those of LLM-based techniques evaluated in RQ1. 
We find that the union of resolved issues further increases to 45.39\%.
These findings reveal remarkably strong complementarity between different LLMs and agents in LLVM issue resolution tasks. 
Although the individual performance of current techniques remains limited, different approaches tend to resolve substantially different subsets of issues, highlighting considerable untapped potential for collaborative or ensemble-based issue resolution systems. 

\begin{findings}
\textbf{\textit{Finding 7:}} 
Different agents exhibit strong complementarity, with each agent resolving distinct subsets of issues.
\end{findings}


\subsection{RQ3: Error Characteristics of LLMs and Agents on \bench{}}
\label{subsec:RQ3}

\subsubsection{Process}
To better understand the limitations of LLMs and agents on \bench{}, we analyze the characteristics of both resolved and unresolved issues from three perspectives: 
(1) \textbf{Task Complexity}: We consider six metrics, including issue description length, the number of modified files, lines, and tokens in the golden patch, human resolution duration, and developer discussion rounds.
(2) \textbf{Task Category}: We analyze issue distributions according to three dimensions, including issue type, LLVM component, and LLVM version.
(3) \textbf{Failure Characteristics of Incorrect Patches}: We analyze incorrect patches across five progressively stricter criteria, including \textit{validity} (the patch can be applied to the LLVM codebase), \textit{compilability} (the patched LLVM codebase can be built), \textit{file-level localization correctness} (the patch modifies the same files as the golden patch), \textit{line-level localization correctness} (the patch modifies the same code lines as the golden patch), and \textit{functional correctness} (the patch passes all LLVM test suites).

\begin{table*}[t!]
    \caption{Distribution of resolved and unresolved instances across three task categories (issue type, component, and version).}
    \label{tab:rq3_categories}
    \centering
    \tabcolsep=1.25mm
    \small
    \begin{tabular}{lccccccccccc}
        \toprule
        \multirow{2}{*}{\textbf{Category}} & \multicolumn{3}{c}{\textbf{Issue Type}} & \multicolumn{4}{c}{\textbf{LLVM Component}} & \multicolumn{4}{c}{\textbf{LLVM Version}} \\ \cmidrule(lr){2-4} \cmidrule(lr){5-8} \cmidrule(lr){9-12} 
        & \textbf{Bug Fix} & \textbf{Opt.} & \textbf{New Feat.} & \textbf{Front-end} & \textbf{Mid-end} & \textbf{Back-end} & \textbf{Others} & \textbf{18.x.x} & \textbf{19.x.x} & \textbf{20.x.x} & \textbf{21.x.x} \\
        \midrule
        \textbf{Resolved Instances} & 3.75\% & 2.41\% & 0.88\% & 3.76\% & 3.12\% & 3.08\% & 3.89\% & 3.28\% & 3.12\% & 2.86\% & 4.23\% \\ [2pt] \hdashline \\[-7pt]
        \textbf{Unresolved Instances} & 96.25\% & 97.59\% & 99.12\% & 96.24\% & 96.88\% & 96.92\% & 96.11\% & 96.72\% & 96.88\% & 97.14\% & 95.77\% \\ 
        \bottomrule
    \end{tabular}
\end{table*}

\begin{table}[t!]
    \caption{Complexity of resolved and unresolved instances.}
    \label{tab:rq3_input_output}
    \centering
    \tabcolsep=1.0mm
    \small
    \begin{tabular}{cccccc}
        \toprule
        \multicolumn{1}{c}{\textbf{Issue}} & \multicolumn{3}{c}{\textbf{Golden Patch}} & \multicolumn{2}{c}{\textbf{Manual Resolution}} \\ \cmidrule(lr){1-1} \cmidrule(lr){2-4} \cmidrule(lr){5-6} 
        \textbf{\#Tokens} & \textbf{\#Files} & \textbf{\#Lines} & \textbf{\#Tokens} & \textbf{Time (day)} & \textbf{\#Discussion} \\
        \midrule
        \multicolumn{6}{l}{\framecolorbox[8.3cm][l]{gray!30}{gray!30}{\textbf{Resolved Instances}}} \\[2pt] \hdashline \\[-7pt]
        1338.47 \:\: & 2.40 \:\: & 59.46 \:\: & 364.75 \:\: & 69.51 \:\: & 2.46 \:\: \\ \midrule
        \multicolumn{6}{l}{\framecolorbox[8.3cm][l]{gray!30}{gray!30}{\textbf{Unresolved Instances}}} \\[2pt] \hdashline \\[-7pt]
        1316.60 \textcolor{red}{$\downarrow$} & 3.39 \textcolor{green}{$\uparrow$} & 132.18 \textcolor{green}{$\uparrow$} & 708.41 \textcolor{green}{$\uparrow$} & 71.76 \textcolor{green}{$\uparrow$} & 3.74 \textcolor{green}{$\uparrow$} \\
        \bottomrule
    \end{tabular}
\end{table}

\subsubsection{Results}
We aggregate the results of all 36 evaluated LLMs and agents and report their overall statistics.

\textbf{Perspective I: Task Complexity.}
Table~\ref{tab:rq3_input_output} compares the task complexity of resolved and unresolved instances. 
First, we observe that unresolved instances have slightly shorter issue descriptions on average than resolved instances. 
Through manual inspection, we find that some resolved instances contain more comprehensive information, such as detailed problem descriptions and reproduction scripts. 
This observation potentially suggests that high-quality and detailed issue reports may facilitate issue resolution and highlights the importance of well-documented issue descriptions in LLVM development and maintenance.
Second, we find that unresolved instances are associated with substantially more complex golden patches than those of resolved instances. 
Specifically, the average numbers of modified files, changed lines, and tokens for unresolved instances are 3.39, 132.18, and 708.41, respectively, compared with 2.40, 59.46, and 364.65 for resolved instances. 
These differences correspond to increases ranging from 29.20\% to 55.02\%. 
These results indicate that unresolved issues generally require larger and more complex code modifications, making them more challenging for both LLMs and agents to solve.
Furthermore, unresolved instances exhibit longer human resolution durations and involve more rounds of developer discussion than resolved instances. 
This finding suggests that the unresolved issues that remain difficult for LLMs and agents are also more challenging for human developers, suggesting a degree of consistency between human and LLM perceptions of issue complexity in LLVM.

\begin{findings}
\textbf{\textit{Finding 8:}} 
Unresolved instances involve larger code modifications, longer human resolution times, and more developer discussion rounds, suggesting that issue complexity is similarly reflected in both human and LLM performance.
\end{findings}

\textbf{Perspective II: Task Category.}
Table~\ref{tab:rq3_categories} presents the distributions of resolved and unresolved instances across three task categories.
For issue types, we first observe resolution rates of 3.72\%, 2.41\%, and 0.88\% for bug-fix, optimization, and new-feature tasks, respectively. 
These results indicate that current LLMs and agents perform substantially better on bug-fix tasks than on new-feature tasks.
A likely reason is that new-feature tasks require significantly more complex code modifications, involving 106.81\% more changed lines on average than bug-fix tasks. 
This observation is consistent with prior studies~\cite{li2025fea,zan2026multi}.
For LLVM components, we observe relatively similar resolution rates across components, ranging from 3.08\% to 3.89\%. 
This result suggests that the specific compiler component involved is not a dominant factor affecting the performance of LLMs and agents.
For LLVM versions, the overall differences across versions are relatively small, indicating that version-specific factors have a limited impact on performance.

\begin{findings}
\textbf{\textit{Finding 9:}} 
Existing techniques perform best on bug-fix issues and worst on new-feature issues, while LLVM components and versions have little impact on performance.
\end{findings}

\begin{table*}[t!]
    \caption{Effectiveness and efficiency comparison of \tech{} and baselines on \bench{}.}
    \label{tab:tech_main}
    \centering
    \tabcolsep=1.2mm
    \small
    \begin{tabular}{llcccccc}
        \toprule
        \multirow{2}{*}{\textbf{\makecell[l]{Candidate\\ Space}}} & \multirow{2}{*}{\textbf{Technique}} & \multicolumn{3}{c}{\textbf{Effectiveness \textcolor{green}{$\uparrow$}}} & \multicolumn{3}{c}{\textbf{Efficiency (AVG.) \textcolor{red}{$\downarrow$}}} \\ \cmidrule(lr){3-5} \cmidrule(lr){6-8}
        & & \textbf{\%Applied} & \textbf{\%Built} & \textbf{\%Resolved} & \textbf{\#Input} & \textbf{\#Output} & \textbf{\$Cost}  \\
        \midrule
        \multirow{4}{*}{\textbf{\makecell[l]{LLMs + Sparse\\(12 Patches)}}} 
        & Individual Method & 17.49$\sim$41.61\% & 2.36$\sim$18.44\% & 0.24$\sim$1.89\% & \cellcolor{red!20}\textbf{14,625$\sim$59,648} & \cellcolor{red!20}\textbf{382$\sim$7,035} & \cellcolor{red!20}\textbf{0.003$\sim$0.101}  \\ 
        & Random Ensemble & 29.31\% & 6.62\% & 0.71\% & 403,049 & 36,724 & 0.306  \\ 
        & \tech{}$_{DeepSeek}$ & \cellcolor{green!20}\textbf{61.94\%} & \cellcolor{green!20}\textbf{54.14\%} & \cellcolor{green!20}\textbf{4.26\%} & 403,220 & 36,750 & 0.306  \\ 
        & \tech{}$_{Grok}$ & \cellcolor{green!20}\textbf{61.94\%} & \cellcolor{green!20}\textbf{54.14\%} & \cellcolor{green!20}\textbf{4.26\%} & 403,135 & 36,811 & 0.306  \\ 
        \midrule
        \multirow{4}{*}{\textbf{\makecell[l]{LLMs + Oracle\\(12 Patches)}}} 
        & Individual Method & 23.88$\sim$53.66\% & 4.73$\sim$35.22\% & 1.42$\sim$4.02\% & \cellcolor{red!20}\textbf{13,602$\sim$42,118} & \cellcolor{red!20}\textbf{399$\sim$7,003} & \cellcolor{red!20}\textbf{0.003$\sim$0.071}  \\ 
        & Random Ensemble & 37.12\% & 20.09\% & 2.84\% & 315,291 & 35,095 & 0.242  \\ 
        & \tech{}$_{DeepSeek}$ & \cellcolor{green!20}\textbf{82.74\%} & \cellcolor{green!20}\textbf{79.43\%} & 12.06\% & 315,711 & 35,170 & 0.242  \\ 
        & \tech{}$_{Grok}$ & \cellcolor{green!20}\textbf{82.74\%} & \cellcolor{green!20}\textbf{79.43\%} & \cellcolor{green!20}\textbf{12.29\%} & 315,691 & 35,345 & 0.242  \\ 
        \midrule 	 	 	
        \multirow{4}{*}{\textbf{\makecell[l]{Agents\\(12 Patches)}}} 
        & Individual Method & 21.75$\sim$55.32\% & 6.86$\sim$40.19\% & 2.60$\sim$10.87\% & \cellcolor{red!20}\textbf{191,273$\sim$500,309} & \cellcolor{red!20}\textbf{5,819$\sim$34,802} & \cellcolor{red!20}\textbf{0.024$\sim$0.601}  \\ 
        & Random Ensemble & 37.35\% & 24.11\% & 7.33\% & 3,520,390 & 168,914 & 2.301  \\ 
        & \tech{}$_{DeepSeek}$ & \cellcolor{green!20}\textbf{90.07\%} & \cellcolor{green!20}\textbf{87.47\%} & 21.04\% & 3,521,301 & 169,101 & 2.301  \\ 
        & \tech{}$_{Grok}$ & \cellcolor{green!20}\textbf{90.07\%} & \cellcolor{green!20}\textbf{87.47\%} & \cellcolor{green!20}\textbf{21.99\%} & 3,521,257 & 169,420 & 2.302  \\  
        \midrule
        \multirow{4}{*}{\textbf{\makecell[l]{LLMs \& Agents\\(36 Patches)}}} 
        & Individual Method & 17.49$\sim$55.32\% & 2.36$\sim$40.19\% & 0.24$\sim$10.87\% & \cellcolor{red!20}\textbf{13,602$\sim$500,309} & \cellcolor{red!20}\textbf{382$\sim$34,802} & \cellcolor{red!20}\textbf{0.003$\sim$0.601}  \\ 
        & Random Ensemble & 33.81\% & 18.68\% & 4.26\% & 4,238,730 & 240,734 & 2.849  \\ 
        & \tech{}$_{DeepSeek}$ & \cellcolor{green!20}\textbf{95.74\%} & \cellcolor{green!20}\textbf{95.51\%} & \cellcolor{green!20}\textbf{20.80\%} & 4,240,303 & 241,016 & 2.849  \\ 
        & \tech{}$_{Grok}$ & \cellcolor{green!20}\textbf{95.74\%} & \cellcolor{green!20}\textbf{95.51\%} & \cellcolor{green!20}\textbf{20.80\%} & 4,240,209 & 241,670 & 2.851  \\ 
        \bottomrule
    \end{tabular}
\end{table*}

\begin{figure}[t!]
    \centering
    \includegraphics[width=1.0\linewidth]{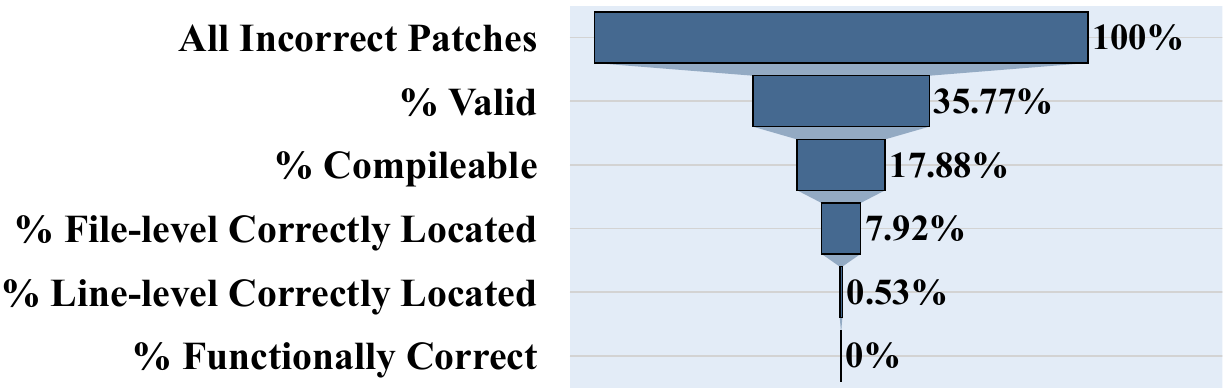}
    \caption{The failure reasons of incorrect patches (funnel chart)}
    \label{fig:rq3_funnel}
\end{figure}

\textbf{Perspective III: Failure Characteristics of Incorrect Patches.}
Figure~\ref{fig:rq3_funnel} presents the distribution of failure reasons for incorrect patches generated by LLMs and agents.
Among all incorrect patches, only 35.77\% of patches can be applied successfully, representing the largest source of failure and highlighting patch validity as a critical challenge for LLVM issue resolution.
Among the incorrect patches that can be successfully applied, only 17.88\% are compilable, indicating that a large proportion of patches introduce program errors and fail at the compiler build stage. 
We further analyze the remaining compilable incorrect patches and find that 9.96\% introduce file-level localization errors, while only 7.39\% introduce line-level localization errors. 
Ultimately, merely 0.53\% of incorrect patches reach the final stage, where they modify the correct code locations but still fail to pass all LLVM tests due to functional errors.
Overall, the primary bottlenecks of current LLMs and agents lie in patch validity and build failures. 
Given that many failures occur before reaching the functional validation stage, these results suggest that current techniques may have difficulty reasoning about repository structures, dependencies, and the constraints necessary to produce compilable patches in large-scale LLVM systems.

\begin{findings}
\textbf{\textit{Finding 10:}} 
Patch invalidity and build failures are the dominant failure modes, occurring before functional validation and suggesting that current techniques struggle to understand the repository structure and dependencies of the large-scale LLVM codebase.
\end{findings}

\subsection{RQ4: Performance of \tech{} on \bench{}}
\label{subsec:RQ4}

\subsubsection{Approach}
Motivated by the strong complementarity among different LLMs and agents (Finding 4 and 7), we propose \textbf{\tech{}}, a lightweight ensemble approach designed to further improve LLVM issue resolution. 
It expands the patch space through integrating the patches generated by multiple techniques, filters incorrect and redundant candidates, and finally identifies the most promising solution.
\begin{itemize}[leftmargin=10pt]
    \item \textbf{Incorrect Patch Filtering}: 
    Given a set of candidate patches, \tech{} first filters incorrect patches using \gym{} (Section~\ref{subsubsec:execution_based_filtering}). 
    Each candidate patch is applied to the LLVM codebase and the resulting compiler is built. 
    Candidate patches that fail to be applied or cause build failures are discarded. 
    This step ensures that only valid and compilable patches are retained for subsequent ensemble.

    \item \textbf{Redundant Patch Filtering}:
    \tech{} then removes duplicate patches through patch normalization and equivalence checking. 
    Following prior work~\cite{tian2026ensemble}, we implement a parser based on \texttt{unidiff}~\cite{unidiff2026} to convert raw patches into standardized representations. 
    Semantically irrelevant elements (e.g., extra spaces, line breaks, and comments) are removed before equivalence checking, and only one representative is retained from each group of equivalent patches.

    \item \textbf{Ensemble Reasoning}:
    Finally, an LLM-based ensemble agent jointly analyzes the issue description and the remaining candidate patches, and then identifies the most promising solution. 
    By leveraging the complementary strengths of diverse LLMs and agents, \tech{} further improves LLVM issue resolution performance.
\end{itemize}

We evaluate the effectiveness and efficiency of \tech{} using the candidate patches generated by LLMs and agents in RQ1 and RQ2. 
In addition, we compare \tech{} against a random ensemble baseline that randomly chooses one patch from all candidate patches.
To investigate the impact of the candidate space, we construct four candidate pools: (1) 12 LLM-based techniques using Sparse Retrieval, (2) 12 LLM-based techniques using Oracle Retrieval, (3) 12 agent-based techniques, and (4) all 36 LLM- and agent-based techniques.
We employ DeepSeek and Grok as the base models for \tech{}. 
For effectiveness evaluation, we report \textit{\%Applied}, \textit{\%Built}, and \textit{\%Resolved}. 
For efficiency evaluation, we report \textit{\#Input Tokens}, \textit{\#Output Tokens}, and \textit{\$Cost}.

\subsubsection{Results}
Table~\ref{tab:tech_main} presents the effectiveness and efficiency of \tech{} on \bench{}.
First, we observe that \tech{} consistently outperforms all individual techniques. 
Compared with individual techniques, \tech{} achieves average improvements of 140.11\%, 361.91\%, and 339.38\% in term of \textit{\%Applied}, \textit{\%Built}, and \textit{\%Resolved}, respectively. 
These results demonstrate that \tech{} further improves the effectiveness of LLVM issue resolution by exploiting the complementary strengths of diverse techniques.
In contrast, the random ensemble baseline performs substantially worse than \tech{}, highlighting that the performance gains of \tech{} are not only attributable to the extended candidate spaces, but also stem from its ability to effectively identify the correct patch from the candidate spaces.

In efficiency evaluation, both \tech{} and the random ensemble baseline incur higher consumption than individual techniques. 
It primarily arises from the extended candidate patches generated by multiple LLMs and agents. 
Nevertheless, given the significant improvements, the additional computational cost of \tech{} is well justified, suggesting a favorable trade-off between effectiveness and efficiency.
Interestingly, utilizing the largest candidate space (i.e., all 36 candidates) does not necessarily yield the highest \textit{\%Resolved} score, suggesting that increasing the candidate space does not monotonically improve ensemble effectiveness. 
While a larger candidate space provides greater diversity, it potentially introduce additional incorrect candidate patches and increase the difficulty of identifying the correct one. 
We plan to further explore the relationship between space size and ensemble effectiveness in future work.

\begin{findings}
\textbf{\textit{Finding 11:}} 
By exploiting the complementarity of diverse techniques, \tech{} achieves an average 339.38\% improvement in \textit{\%Resolved} over each individual technique, albeit at higher computational costs.
\end{findings}
\section{Implications}
\label{sec:implications}

Our findings suggest several implications for future research on large-scale system software such as LLVM.

\noindent
\ul{\textbf{Implication 1: Benchmarking LLMs on system software.}}
Existing issue resolution benchmarks primarily focus on application-level software, whereas LLVM represents a significantly more challenging system software with a massive codebase, complex dependencies, and a rigorous testing pipeline.
Future evaluations should extend beyond application software to large-scale system software.
We hope that \bench{} and \gym{} will provide a standardized and extensible infrastructure for benchmarking and advancing LLMs and agents on large-scale system software.

\noindent
\ul{\textbf{Implication 2: Repository context should be summarized, not merely expanded.}}
Although larger retrieval contexts consistently improve issue resolution performance, they also incur substantially higher computational cost. 
Rather than merely increasing context length, future work should develop context summarization and compression techniques that preserve critical issue-relevant information while eliminating redundancy, thereby improving both effectiveness and efficiency.

\noindent
\ul{\textbf{Implication 3: Complementarity-aware ensemble method.}}
Our study demonstrates that exploiting the complementarity of diverse techniques substantially improves LLVM issue resolution, but generating candidate patches from many techniques is computationally expensive.
Future ensemble methods should explicitly analyze complementarity and identify a small yet diverse set of techniques that maximizes coverage while minimizing redundant candidate patches.

\noindent
\ul{\textbf{Implication 4: Task-aware agent workflows.}}
New-feature tasks are considerably more difficult than bug-fix tasks, suggesting that a unified workflow (e.g., reproduction, localization, and repair) is insufficient for all issue types. 
Future agents should adopt task-aware workflows. 
For example, new-feature resolution may require dedicated stages for requirement generation, feature implementation, and functionality validation, rather than relying solely on bug-fixing pipelines.

\noindent
\ul{\textbf{Implication 5: Static and dynamic feedback for patch generation.}}
Patch invalidity and build failures are the dominant failure modes, indicating that current techniques struggle to satisfy LLVM-specific implementation constraints. 
Future work should combine static analysis (e.g., ClangFormat~\cite{clang2026clangformat} and Clang-Tidy~\cite{clang2026clangtidy}) with dynamic feedback from \gym{} (e.g., patch application and compiler building) to support iterative patch generation and refinement.

\section{Discussion}
\label{sec:discussion}


\subsection{Future Applicability to Emerging LLMs and Agents}
We acknowledge that newly released LLMs and agents may achieve higher absolute performance than the LLMs and agents evaluated in this work. 
Nevertheless, we believe our main conclusions are unlikely to change. 
\uline{Recent results show that the state-of-the-art combination of Claude-Code~\mbox{\cite{anthropic_claude_code}} and Claude-Opus-4.8~\mbox{\cite{anthropic2026opus48}} achieves 85.8\% \textit{\%Resolved} on SWE-bench-Verified~\mbox{\cite{swebenchvirified2024}}, whereas the best configuration evaluated in this work (SWE-agent with Gemini-3-Flash) achieves 78\% on the same benchmark but only 4.26\% on \bench{}.}
This substantial performance gap suggests that LLVM issue resolution remains significantly more challenging than existing software engineering benchmarks, even for increasingly capable models. 
To facilitate future evaluation, we will maintain a public leaderboard and continuously update \bench{} with new LLVM issues.

\subsection{Generalizability to Other Large-Scale System Software}
Although this study focuses on LLVM, we believe that many of our findings to generalize to other large-scale system software, such as GCC~\cite{gough2004introduction}, PostgreSQL~\cite{douglas2003postgresql}, and Kubernetes~\cite{burns2022kubernetes}. 
These systems share several key characteristics with LLVM, including massive codebases, complex dependencies, and rigorous test suites, making issue resolution fundamentally more challenging than in the existing application-level benchmarks. 
Consequently, the findings in this work, including the limited effectiveness of current issue resolution techniques and the complementarity among different LLMs and agents, are likely to extend beyond LLVM. 
Future work includes applying our benchmark construction and evaluation methodology to other large-scale system software to validate these findings.

\subsection{Threats and Validity}
\label{sec:threats}

\noindent
\textbf{Internal Validity.} 
To reduce the randomness of LLM-based techniques, all methods are evaluated on the full set of 423 benchmark instances with the temperature fixed at 0. In addition, all manual validation is conducted independently by multiple authors and finalized through consensus.

\noindent
\textbf{External Validity.} 
The primary threat concerns the generalizability of our findings. 
To mitigate this threat, \bench{} is constructed from real LLVM issues and merged pull requests, ensuring that it reflects authentic development scenarios. 
Moreover, we evaluate four representative LLMs, six retrieval settings, and three widely used agents, providing broad coverage of current issue resolution techniques.

\noindent
\textbf{Construct Validity.} 
A potential threat is whether our evaluation metrics adequately capture issue resolution performance. 
To address this, we measure effectiveness using \textit{\%Applied}, \textit{\%Built}, and \textit{\%Resolved}, and efficiency using \textit{\#Input Tokens}, \textit{\#Output Tokens}, and \textit{\$Cost}, which evaluate validity, compilability, and functional correctness, respectively.
\section{Related Work}
\label{sec:related}

Automatic software issue resolution has become a fundamental software engineering task, motivating the development of numerous benchmarks for evaluating LLM-based issue resolution techniques.
The first and most influential benchmark, SWE-bench~\cite{jimenezswe}, contains real-world issue resolution tasks collected from 12 popular Python projects. 
To improve evaluation reliability, SWE-bench-Verified~\cite{swebenchvirified2024} provides a manually verified subset. 
More recently, researchers have extended issue resolution benchmarks to broader software ecosystems, including Rust-SWE-bench~\cite{xiang2026evaluating}, SWE-PolyBench~\cite{rashid2025swe}, Multi-SWE-bench~\cite{zan2026multi}, SWE-bench-Multilingual~\cite{yang2026swesmith}, SWE-bench-Live~\cite{zhang2026swe}, and SWE-bench Pro~\cite{deng2025swepro}, covering multiple programming languages, repositories, and increasingly challenging issue resolution tasks. 
Beyond application software, kBenchSyz~\cite{mathai2024kgym} and LivekBench~\cite{huang2026outrunning} introduce benchmarks for Linux kernel crash resolution from Syzbot~\cite{google2026syzbot}.

Despite this progress, existing benchmarks primarily target application software or operating systems and do not cover large-scale compiler infrastructures such as LLVM. 
Consequently, the effectiveness of current LLM-based issue resolution techniques on complex compiler development tasks remains largely unexplored. 
To bridge this gap, we introduce \bench{}, the first large-scale benchmark for LLVM issue resolution, enabling systematic evaluation of LLM-based issue resolution techniques on LLVM compiler.

\section{Conclusion}
This paper presents \bench{}, the first large-scale benchmark for real-world LLVM issue resolution, comprising 423 high-quality, validated tasks. 
We further develop \gym{}, a scalable evaluation platform that automates issue reproduction, patch application, compiler building, and LLVM test execution.
Using \bench{} and \gym{}, we conduct a comprehensive evaluation of LLM-based issue resolution techniques and find that current techniques remain limited on \bench{}. 
Motivated by the strong complementarity among different LLMs and agents, we propose \tech{}, a lightweight ensemble approach that expands the patch space through integrating the patches generated by diverse techniques, filters incorrect and redundant candidates, and identifies the most promising solution.
Experimental results demonstrate that \tech{} further improves the effectiveness of LLVM issue resolution.


\balance
\bibliographystyle{IEEEtran}
\bibliography{reference}

@article{zhang2026swe,
  title={Swe-bench goes live!},
  author={Zhang, Linghao and He, Shilin and Zhang, Chaoyun and Kang, Yu and Li, Bowen and Xie, Chengxing and Wang, Junhao and Wang, Maoquan and Huang, Yufan and Fu, Shengyu and others},
  journal={Advances in Neural Information Processing Systems},
  volume={38},
  year={2026}
}

@misc{clang2026clangtidy,
  author       = {The Clang-Tidy Team},
  year         = {2026},
  title        = {Clang-Tidy},
  howpublished = {\url{https://clang.llvm.org/extra/clang-tidy/}},
}

@misc{clang2026clangformat,
  author       = {The ClangFormat Team},
  year         = {2026},
  title        = {ClangFormat},
  howpublished = {\url{https://clang.llvm.org/docs/ClangFormat.html}},
}

@misc{google2026syzbot,
  author       = {Google},
  year         = {2026},
  title        = {Syzbot},
  howpublished = {\url{https://syzkaller.appspot.com/upstream}},
}

@article{huang2026outrunning,
  title={Outrunning LLM Cutoffs: A Live Kernel Crash Resolution Benchmark for All},
  author={Huang, Chenxi and Mathai, Alex and Yu, Feiyang and Nogikh, Aleksandr and Maniatis, Petros and Ivan{\v{c}}i{\'c}, Franjo and Wu, Eugene and Kaffes, Kostis and Yang, Junfeng and Ray, Baishakhi},
  journal={arXiv preprint arXiv:2602.02690},
  year={2026}
}

@article{mathai2024kgym,
  title={Kgym: A platform and dataset to benchmark large language models on linux kernel crash resolution},
  author={Mathai, Alex and Huang, Chenxi and Maniatis, Petros and Nogikh, Aleksandr and Ivan{\v{c}}i{\'c}, Franjo and Yang, Junfeng and Ray, Baishakhi},
  journal={Advances in Neural Information Processing Systems},
  volume={37},
  pages={78053--78078},
  year={2024}
}

@book{burns2022kubernetes,
  title={Kubernetes: up and running: dive into the future of infrastructure},
  author={Burns, Brendan and Beda, Joe and Hightower, Kelsey and Evenson, Lachlan},
  year={2022},
  publisher={" O'Reilly Media, Inc."}
}

@book{douglas2003postgresql,
  title={PostgreSQL: a comprehensive guide to building, programming, and administering PostgresSQL databases},
  author={Douglas, Korry and Douglas, Susan},
  year={2003},
  publisher={SAMS publishing}
}

@misc{anthropic2026opus48,
  author       = {Anthropic},
  year         = {2026},
  title        = {Introducing Claude Opus 4.8},
  howpublished = {\url{https://www.anthropic.com/news/claude-opus-4-8}},
}

@misc{anthropic_claude_code,
  author       = {The Claude Code Team},
  year         = {2026},
  title        = {Claude Code: AI-powered coding assistant for developers},
  howpublished = {\url{https://www.anthropic.com/claude-code}},
}

@book{gough2004introduction,
  title={An Introduction to GCC.},
  author={Gough, Brian J and Stallman, Richard},
  year={2004},
  publisher={Network Theory Limited Bristol, UK}
}

@inproceedings{robertson1994some,
  title={Some simple effective approximations to the 2-poisson model for probabilistic weighted retrieval},
  author={Robertson, Stephen E and Walker, Steve},
  booktitle={SIGIR’94: Proceedings of the Seventeenth Annual International ACM-SIGIR Conference on Research and Development in Information Retrieval, organised by Dublin City University},
  pages={232--241},
  year={1994},
  organization={Springer}
}

@inproceedings{ruan2025specrover,
  title={Specrover: Code intent extraction via llms},
  author={Ruan, Haifeng and Zhang, Yuntong and Roychoudhury, Abhik},
  booktitle={2025 IEEE/ACM 47th International Conference on Software Engineering (ICSE)},
  pages={963--974},
  year={2025},
  organization={IEEE}
}

@article{ma2025swe,
  title={Swe-gpt: A process-centric language model for automated software improvement},
  author={Ma, Yingwei and Cao, Rongyu and Cao, Yongchang and Zhang, Yue and Chen, Jue and Liu, Yibo and Liu, Yuchen and Li, Binhua and Huang, Fei and Li, Yongbin},
  journal={Proceedings of the ACM on Software Engineering},
  volume={2},
  number={ISSTA},
  pages={2362--2383},
  year={2025},
  publisher={ACM New York, NY, USA}
}

@article{hou2024large,
  title={Large language models for software engineering: A systematic literature review},
  author={Hou, Xinyi and Zhao, Yanjie and Liu, Yue and Yang, Zhou and Wang, Kailong and Li, Li and Luo, Xiapu and Lo, David and Grundy, John and Wang, Haoyu},
  journal={ACM Transactions on Software Engineering and Methodology},
  volume={33},
  number={8},
  pages={1--79},
  year={2024},
  publisher={ACM New York, NY}
}

@article{zhou2021empirical,
  title={An empirical study of optimization bugs in GCC and LLVM},
  author={Zhou, Zhide and Ren, Zhilei and Gao, Guojun and Jiang, He},
  journal={Journal of Systems and Software},
  volume={174},
  pages={110884},
  year={2021},
  publisher={Elsevier}
}

@article{wang2025solved,
  title={Are" solved issues" in swe-bench really solved correctly? an empirical study},
  author={Wang, You and Pradel, Michael and Liu, Zhongxin},
  journal={2026 IEEE/ACM 48th International Conference on Software Engineering (ICSE)},
  year={2026}
}

@article{li2025swe,
  title={Swe-debate: Competitive multi-agent debate for software issue resolution},
  author={Li, Han and Shi, Yuling and Lin, Shaoxin and Gu, Xiaodong and Lian, Heng and Wang, Xin and Jia, Yantao and Huang, Tao and Wang, Qianxiang},
  journal={2026 IEEE/ACM 48th International Conference on Software Engineering (ICSE)},
  year={2026}
}

@article{deng2025swepro,
  title={Swe-bench pro: Can ai agents solve long-horizon software engineering tasks?},
  author={Deng, Xiang and Da, Jeff and Pan, Edwin and He, Yannis Yiming and Ide, Charles and Garg, Kanak and Lauffer, Niklas and Park, Andrew and Pasari, Nitin and Rane, Chetan and others},
  journal={arXiv preprint arXiv:2509.16941},
  year={2025}
}

@article{yang2026swesmith,
  title={Swe-smith: Scaling data for software engineering agents},
  author={Yang, John and Lieret, Kilian and Jimenez, Carlos and Wettig, Alexander and Khandpur, Kabir and Zhang, Yanzhe and Hui, Binyuan and Press, Ofir and Schmidt, Ludwig and Yang, Diyi},
  journal={Advances in Neural Information Processing Systems},
  volume={38},
  year={2026}
}

@article{rashid2025swe,
  title={Swe-polybench: A multi-language benchmark for repository level evaluation of coding agents},
  author={Rashid, Muhammad Shihab and Bock, Christian and Zhuang, Yuan and Buchholz, Alexander and Esler, Tim and Valentin, Simon and Franceschi, Luca and Wistuba, Martin and Sivaprasad, Prabhu Teja and Kim, Woo Jung and others},
  journal={arXiv preprint arXiv:2504.08703},
  year={2025}
}

@article{tian2026ensemble,
  title={Agent-Based Ensemble Reasoning for Repository-Level Issue Resolution},
  author={Tian, Zhao and Gao, Pengfei and Chen, Junjie and Peng, Chao},
  journal={2026 IEEE/ACM 48th International Conference on Software Engineering (ICSE)},
  year={2026}
}

@article{li2025fea,
  title={Fea-bench: A benchmark for evaluating repository-level code generation for feature implementation},
  author={Li, Wei and Zhang, Xin and Guo, Zhongxin and Mao, Shaoguang and Luo, Wen and Peng, Guangyue and Huang, Yangyu and Wang, Houfeng and Li, Scarlett},
  journal={Proceedings of the 63rd Annual Meeting of the Association for Computational Linguistics (Volume 1: Long Papers)},
  pages={17160--17176},
  year={2025}
}

@article{zan2026multi,
  title={Multi-swe-bench: A multilingual benchmark for issue resolving},
  author={Zan, Daoguang and Huang, Zhirong and Liu, Wei and Chen, Hanwu and Xin, Shulin and Zhang, Linhao and Liu, Qi and Aoyan, Li and Chen, Lu and Zhong, Xiaojian and others},
  journal={Advances in Neural Information Processing Systems},
  volume={38},
  year={2026}
}

@article{xiang2026evaluating,
  title={Evaluating and Improving Automated Repository-Level Rust Issue Resolution with LLM-based Agents},
  author={Xiang, Jiahong and He, Wenxiao and Wang, Xihua and Tian, Hongliang and Zhang, Yuqun},
  journal={2026 IEEE/ACM 48th International Conference on Software Engineering (ICSE)},
  year={2026}
}

@misc{ninja2026,
  year = {2026},
  title = {Ninja, a small build system with a focus on speed},
  author={Ninja},
  howpublished ={\url{https://ninja-build.org/}}
}

@article{xia2025live,
  title={Live-SWE-agent: Can Software Engineering Agents Self-Evolve on the Fly?},
  author={Xia, Chunqiu Steven and Wang, Zhe and Yang, Yan and Wei, Yuxiang and Zhang, Lingming},
  journal={arXiv preprint arXiv:2511.13646},
  year={2025}
}

@article{yang2025qwen3,
  title={Qwen3 technical report},
  author={Yang, An and Li, Anfeng and Yang, Baosong and Zhang, Beichen and Hui, Binyuan and Zheng, Bo and Yu, Bowen and Gao, Chang and Huang, Chengen and Lv, Chenxu and others},
  journal={arXiv preprint arXiv:2505.09388},
  year={2025}
}

@article{liu2024deepseek,
  title={Deepseek-v3 technical report},
  author={Liu, Aixin and Feng, Bei and Xue, Bing and Wang, Bingxuan and Wu, Bochao and Lu, Chengda and Zhao, Chenggang and Deng, Chengqi and Zhang, Chenyu and Ruan, Chong and others},
  journal={arXiv preprint arXiv:2412.19437},
  year={2024}
}

@misc{grok2026,
  year = {2026},
  title = {Grok},
  author={xAI},
  howpublished ={\url{https://x.ai/grok}}
}

@article{team2023gemini,
  title={Gemini: a family of highly capable multimodal models},
  author={Team, Gemini and Anil, Rohan and Borgeaud, Sebastian and Alayrac, Jean-Baptiste and Yu, Jiahui and Soricut, Radu and Schalkwyk, Johan and Dai, Andrew M and Hauth, Anja and Millican, Katie and others},
  journal={arXiv preprint arXiv:2312.11805},
  year={2023}
}

@misc{llvmbugreport2026,
  year = {2026},
  title = {How to submit an LLVM bug report},
  author={LLVM},
  howpublished ={\url{https://llvm.org/docs/HowToSubmitABug.html}}
}

@misc{pytest2026,
  year = {2026},
  title = {Pytest: helps you write better programs},
  author={Pytest-dev Team},
  howpublished ={\url{https://docs.pytest.org/en/stable/}}
}

@misc{django2026,
  year = {2026},
  title = {Django: The web framework for perfectionists with deadlines.},
  author={Django Software Foundation},
  howpublished ={\url{https://www.djangoproject.com/}}
}

@misc{valsai2026,
  year = {2026},
  title = {Leaderboard of SWE-bench-Verified},
  author={Vals AI},
  howpublished ={\url{https://www.vals.ai/benchmarks/swebench}}
}

@misc{llvmissues2026,
  year = {2026},
  title = {Issues of LLVM Project},
  author={LLVM},
  howpublished ={\url{https://github.com/llvm/llvm-project/issues}}
}

@article{junod2015obfuscator,
  title={Obfuscator-LLVM--software protection for the masses},
  author={Junod, Pascal and Rinaldini, Julien and Wehrli, Johan and Michielin, Julie},
  journal={2015 ieee/acm 1st international workshop on software protection},
  pages={3--9},
  year={2015},
  organization={IEEE}
}

@article{lam2015numba,
  title={Numba: A llvm-based python jit compiler},
  author={Lam, Siu Kwan and Pitrou, Antoine and Seibert, Stanley},
  journal={Proceedings of the Second Workshop on the LLVM Compiler Infrastructure in HPC},
  pages={1--6},
  year={2015}
}

@article{chen2020survey,
  title={A survey of compiler testing},
  author={Chen, Junjie and Patra, Jibesh and Pradel, Michael and Xiong, Yingfei and Zhang, Hongyu and Hao, Dan and Zhang, Lu},
  journal={Acm Computing Surveys (Csur)},
  volume={53},
  number={1},
  pages={1--36},
  year={2020},
  publisher={ACM New York, NY, USA}
}

@article{lattner2004llvm,
  title={LLVM: A compilation framework for lifelong program analysis \& transformation},
  author={Lattner, Chris and Adve, Vikram},
  journal={International symposium on code generation and optimization, 2004. CGO 2004.},
  pages={75--86},
  year={2004},
  organization={IEEE}
}

@article{sun2016toward,
  title={Toward understanding compiler bugs in GCC and LLVM},
  author={Sun, Chengnian and Le, Vu and Zhang, Qirun and Su, Zhendong},
  journal={Proceedings of the 25th international symposium on software testing and analysis},
  pages={294--305},
  year={2016}
}

@article{gao2025trae,
  title={Trae agent: An llm-based agent for software engineering with test-time scaling},
  author={Gao, Pengfei and Tian, Zhao and Meng, Xiangxin and Wang, Xinchen and Hu, Ruida and Xiao, Yuanan and Liu, Yizhou and Zhang, Zhao and Chen, Junjie and Gao, Cuiyun and others},
  journal={arXiv preprint arXiv:2507.23370},
  year={2025}
}

@misc{unidiff2026,
  year = {2026},
  title = {Unified diff python parsing/metadata extraction library},
  author={Matias Bordese},
  howpublished ={\url{https://github.com/matiasb/python-unidiff}}
}

@article{
    wang2025openhands,
    title={OpenHands: An Open Platform for {AI} Software Developers as Generalist Agents},
    author={Xingyao Wang and Boxuan Li and Yufan Song and Frank F. Xu and Xiangru Tang and Mingchen Zhuge and Jiayi Pan and Yueqi Song and Bowen Li and Jaskirat Singh and Hoang H. Tran and Fuqiang Li and Ren Ma and Mingzhang Zheng and Bill Qian and Yanjun Shao and Niklas Muennighoff and Yizhe Zhang and Binyuan Hui and Junyang Lin and Robert Brennan and Hao Peng and Heng Ji and Graham Neubig},
    journal={The Thirteenth International Conference on Learning Representations},
    year={2025}
}

@article{yang2024swe,
  title={Swe-agent: Agent-computer interfaces enable automated software engineering},
  author={Yang, John and Jimenez, Carlos E and Wettig, Alexander and Lieret, Kilian and Yao, Shunyu and Narasimhan, Karthik and Press, Ofir},
  journal={Advances in Neural Information Processing Systems},
  volume={37},
  pages={50528--50652},
  year={2024}
}

@article{xia2024agentless,
    title = {Demystifying LLM-Based Software Engineering Agents},
    author = {Xia, Chunqiu Steven and Deng, Yinlin and Dunn, Soren and Zhang, Lingming},
    year = {2025},
    journal = {Proc. ACM Softw. Eng.},
    volume = {2},
    number = {FSE},
    numpages = {24}
}

@misc{swebenchvirified2024,
  year = {2026},
  title = {Introducing SWE-bench Verified},
  author={OpenAI},
  howpublished ={\url{https://openai.com/index/introducing-swe-bench-verified/}}
}

@article{jimenezswe,
  title={SWE-bench: Can Language Models Resolve Real-world Github Issues?},
  author={Jimenez, Carlos E and Yang, John and Wettig, Alexander and Yao, Shunyu and Pei, Kexin and Press, Ofir and Narasimhan, Karthik R},
  journal={The Twelfth International Conference on Learning Representations},
  year={2023}
}

@misc{clang2026,
  year = {2026},
  title = {Clang: a C language family frontend for LLVM},
  author={Clang},
  howpublished ={\url{https://clang.llvm.org/}}
}

@misc{flang2026,
  year = {2026},
  title = {The Flang Compiler},
  author={Flang},
  howpublished ={\url{https://flang.llvm.org/docs/}}
}

@misc{mlir2026,
  year = {2026},
  title = {Multi-Level IR Compiler Framework},
  author={MLIR},
  howpublished ={\url{https://mlir.llvm.org/}}
}

@misc{rust2026,
  year = {2026},
  title = {Rust Compiler Development Guide},
  author={Rustc},
  howpublished ={\url{https://rustc-dev-guide.rust-lang.org/backend/codegen.html}}
}

@misc{retdec2026,
  year = {2026},
  title = {The RetDec Decompiler},
  author={RetDec},
  howpublished ={\url{https://github.com/avast/retdec}}
}

@inproceedings{DBLP:conf/osdi/CadarDE08,
  author       = {Cristian Cadar and
                  Daniel Dunbar and
                  Dawson R. Engler},
  editor       = {Richard Draves and
                  Robbert van Renesse},
  title        = {{KLEE:} Unassisted and Automatic Generation of High-Coverage Tests
                  for Complex Systems Programs},
  booktitle    = {8th {USENIX} Symposium on Operating Systems Design and Implementation,
                  {OSDI} 2008, December 8-10, 2008, San Diego, California, USA, Proceedings},
  pages        = {209--224},
  publisher    = {{USENIX} Association},
  year         = {2008},
  url          = {http://www.usenix.org/events/osdi08/tech/full\_papers/cadar/cadar.pdf},
  bibsource    = {dblp computer science bibliography, https://dblp.org}
}

@inproceedings{DBLP:conf/tacas/SchubertHB19,
  author       = {Philipp Dominik Schubert and
                  Ben Hermann and
                  Eric Bodden},
  editor       = {Tom{\'{a}}s Vojnar and
                  Lijun Zhang},
  title        = {PhASAR: An Inter-procedural Static Analysis Framework for {C/C++}},
  booktitle    = {Tools and Algorithms for the Construction and Analysis of Systems
                  - 25th International Conference, {TACAS} 2019, Held as Part of the
                  European Joint Conferences on Theory and Practice of Software, {ETAPS}
                  2019, Prague, Czech Republic, April 6-11, 2019, Proceedings, Part
                  {II}},
  series       = {Lecture Notes in Computer Science},
  pages        = {393--410},
  publisher    = {Springer},
  year         = {2019},
  url          = {https://doi.org/10.1007/978-3-030-17465-1\_22},
  doi          = {10.1007/978-3-030-17465-1\_22},
  bibsource    = {dblp computer science bibliography, https://dblp.org}
}

\end{document}